# Examining the Interface Design of Tidyverse

Emi Tanaka

The `tidyverse` is a popular meta-package comprising several core R packages to aid in various data science tasks, including data import, manipulation and visualisation. Although functionalities offered by the `tidyverse` can generally be replicated using other packages, its widespread adoption in both teaching and practice indicates there are factors contributing to its preference, despite some debate over its usage. This suggests that particular aspects, such as interface design, may play a significant role in its selection. Examining the interface design can potentially reveal aspects that aid the design process for developers. While Tidyverse has been lauded for adopting a user-centered design, arguably some elements of the design focus on the work domain instead of the end-user. We examine the Tidyverse interface design via the lens of human computer interaction, with an emphasis on data visualisation and data wrangling, to identify factors that might serve as a model for developers designing their packages. We recommend that developers adopt an iterative design that is informed by user feedback, analysis and complete coverage of the work domain, and ensure perceptual visibility of system constraints and relationships.

## 1 Introduction

The use of specialized programming tools is fundamental in data analysis; however, limited attention has been given to their design beyond select discussions in the literature (e.g. N. S. Matloff 2012; Wickham 2024). The significance of design in statistical software is widely acknowledged, as exemplified in the criteria for the John M. Chambers Statistical Software Award, which prioritizes submissions "grounded in software design rather than calculation" (American Statistical Association 2024). Notably, interface design of analytical tools serves as a communication channel between developers and analysts. Furthermore, when the analysts' actions are documented in the form of code, the interface also facilitates communication between the authors and readers. Advancing best practice and principles in the interface design of analytical tools can promote improved analytical workflows and foster more effective communication (Parker 2017).

To begin, we study what works in widely adopted analytical tools, noting that the terms "analytics" and "statistics" are used interchangeably throughout this paper. R (R Core Team 2024) is regarded as one of the foremost languages for statistics, as evidenced by various metrics (Muenchen 2013; Lai et al. 2019). This is largely due to its extensive capabilities, supported by over 21,000 contributed R packages available on the Comprehensive R Archive Network (CRAN). In particular, `tidyverse` (Wickham et al. 2019) has emerged as one of the most popular collection of contributed R packages that aid in data import, cleaning, wrangling and visualisation tasks (refer to Supplementary materials for empirical evidence). Note that we make a distinction between `tidyverse` and Tidyverse – when we use the lowercase monospace font, `tidyverse` refers to the R meta-package that loads the nine core packages: `dplyr`, `forcats`, `ggplot2`, `lubridate`, `purrr`, `readr`, `stringr`, `tibble`, and `tidyr` (see Supplementary materials for the purpose of each package), while the Tidyverse (in title case, with regular font) contains more than just these nine core packages (for a full list, see the [Tidyverse website](#)).

Despite a number of functional purposes in `tidyverse` being already available in the vanilla installation of R (referred henceforth as Base R), `tidyverse` has gained widespread adoption in practice and teaching (Çetinkaya-Rundel et al. 2022; Staples 2023). For some functions, the primary difference between the approaches in Tidyverse and Base R are in the interface design and there are no noticeable difference in performance for moderately sized data. For example, `filter()` and `select()` are functions in a core package of Tidyverse that can subset the data by row and by column, respectively, but this can also be achieved by `subset()` in Base R (see an extensive comparison in Tanaka 2025). This observation suggests that the interface design could be a crucial factor contributing to the Tidyverse's success.

Several aspects of the Tidyverse have been explored in the literature. For instance, Staples (2023) documents the increased use of Tidyverse based on an analysis of code from public GitHub repositories. Meanwhile, Çetinkaya-Rundel et al. (2022) emphasizes the pedagogical advantages of the Tidyverse, such as its the user-centered design (UCD), and the consistency and readability of its function interfaces. However, McNamara (2024) reports no conclusive benefit using Tidyverse over the formula syntax (and vice versa) in teaching modelling in introductory statistics, although students were observed to spend more time on average computing when using Tidyverse. The pedagogical benefits outlined by Çetinkaya-Rundel et al. (2022) align closely with the design principles of the Tidyverse, as summarized by Wickham (2024): human-centered, consistency, composability and inclusivity. However, these attributes do not clearly explain how other developers can design their own packages. In this paper, we aim to identify factors that may serve as a model for developers designing their packages.

To analyse the interface design of Tidyverse, we adopt a cross-disciplinary perspective from the field of human-computer interaction (HCI). The concept of UCD emerged within HCI as a design framework that focus on addressing user needs throughout the product development process (Norman and Draper 1986). UCD specifically highlights the significance of interface design as a bridge between users and the program or environment, incorporating user feedback, and considering human cognitive processes in interface development. This focus on human cognition led to the emergence of a distinct field known as cognitive ergonomics or cognitive (systems) engineering. In relation to cognitive ergonomics, other interface design approaches have also emerged. Although the Tidyverse design is often characterized as UCD, there are some debatable aspects to this characterisation. This suggests a mixture of design approaches are used, hence other developers may benefit from using this mixed approach. We elaborate on additional perspectives later in Section 3.

This paper is structured as follows. Section 2 provides a comparison of the interface design of Tidyverse with other approaches for common tasks for data visualisation and wrangling. The principles and methodologies of user interface design are outlined in Section 3, with their relevance to Tidyverse discussed in Section 4. Finally, the paper concludes with a discussion and some recommendations in Section 5.

## 2 Comparison of interface designs for data visualisation and wrangling

The Tidyverse encompasses a collection of R packages that share a unified design philosophy, grammar, and data structures to facilitate data science tasks (Wickham, Çetinkaya-Rundel, and Grolemund 2023). The Tidyverse has inspired other packages that adopt similar design principles as explained in Section 2.1.

To illustrate the Tidyverse syntax, we offer comparisons with Base R and other well-known R packages, specifically, `lattice` and `data.table`, for common tasks in data visualisation (Section 2.2) and data wrangling (Section 2.3), using the `penguins` data in the `palmerpenguins` package (Horst, Hill, and Gorman 2022). This data comprises observations of 344 penguins with a mix of numerical and categorical variables (e.g. sex, species, bill length and bill depth).

```
data(penguins, package = "palmerpenguins")
```

A more comprehensive comparison of data wrangling syntax is available in Tanaka (2025). For additional comparisons, refer to the works of Çetinkaya-Rundel et al. (2022), McNamara (2024), Stoudt (2024), and the vignette in Wickham et al. (2023).

## 2.1 Tidy approach

The Tidyverse has evolved into a dialect within the R programming language (Staples 2023), with numerous packages and programming styles adopting the so-called "tidy" approach (Wang, Cook, and Hyndman 2020; Carpenter et al. 2021; Couch et al. 2021; Li, Deans, and Buell 2023; Hernangómez 2023; Shen and Snyder 2023; Zhang et al. 2024a, 2024b; Pedersen 2024; Hutchison et al. 2024). The tidy approach generally embodies the spirit outlined in the design principles by Wickham (2024), which includes human-centered design, consistency, composability, and inclusivity. Typically, the tidy approach exhibits several key characteristics:

- It uses one or more of the core Tidyverse packages under the hood.
- Functions and arguments follow a logical pattern or human-readable naming system.
- There is consistency in the type of input data and the resulting output object.
- Each function is generally designed to perform a single specific purpose, with user-friendly messages when system errors occur or when input/output objects do not meet expectations.
- The primary output is generated by combining modular functions, rather than relying on a single function with numerous arguments.
- The first argument is often the data itself, facilitating sequential processing using pipes `%>%` from the `magrittr` package (Bache and Wickham 2022), or alternatively, the native pipe `|>` available from R version 4.1.0 onwards.
- When dealing with tabular data, the data is expected to adhere to or extend the tidy data principles, where each column represents a variable, each row represents an observation, and each cell contains a value (Wickham 2014).
- It adopts one or more bespoke paradigms within the Tidyverse in Table 1.

Table 1: Bespoke paradigms within the Tidyverse with examples

| Paradigms | Examples |
| --- | --- |
| - Vector prototyping via `vctrs` | Code 2.1 Line 3 |

| Paradigms | Examples |
| --- | --- |
| • Tidy evaluation or data-masking via `rlang` | Code 2.1 Lines 9-11 |
| • `pillar` for supporting aesthetically pleasing formatting in `tibble` | Code 2.1 Output from Line 14 |
| • `purrr`-like formula for lambda functions | Code 2.1 Line 14 argument |
| • Selection helpers for variables from a list via `tidyselect` | See vignette of `tidyselect` |
| • Pretty message styling with `cli` | See vignette of `cli` |

Code 2.1 demonstrates some of the Tidyverse paradigms in Table 1. These demonstrations only show snippets of Tidyverse functionalities. Readers are encouraged to read the vignettes of the corresponding package for more details.

For Code 2.1, line 1 stores the `demo` data based on the first data point in the `penguins` data. Lines 3-5 sets up a function, `labelled`, to create a new class of vector ("label") prototyped with `vctrs` designed to hold the unit of measurement as an attribute "unit". The benefit of prototyped using `vctrs` is that there are various modular S3 generic functions that can be customised to change specific component of the output. For example, line 6 sets up a method (`vec_ptype_abbr.label`) that define the abbreviated type of the vector class "label" as the attribute "unit". This method is called in a number of places, such as deep within `pillar_shaft()` S3 generic function in `pillar` to print the abbreviated type as explained more later. In lines 8-12, we define a function designed to add a unit of measurement to a column in the tabular data. Line 9 transforms the input expression `x` into a string and is similar to `deparse(substitute(x))` in Base R, except it does not necessitate `x` to be an unquoted expression. Line 10 transforms the formula syntax to a function where specially reserved symbols are used for argument placeholders; in this example, the dot `.` is replaced with the first argument. Line 11 then uses `mutate()` from the `dplyr` package to evaluate the expressions (via `rlang` in the backend) with data-masking. More specifically, data-masking in the Tidyverse paradigm treats the named elements of the input data as existing objects in an environment and uses specially reserved pronouns `.data` and `.env` to disambiguate between environments (see more in the vignette about data-masking in `rlang` package). Line 14 calls on the function to add the unit of measurement "cm" to `bill_depth_mm` after dividing it by 100. The `tibble` output from line 14 shows just under `bill_depth_mm`, the unit of measurement is printed as "<cm>". This print is cascaded from the abbreviated type `vec_ptype_abbr.label` to the print by `pillar_shaft()` in the `pillar` package.

**Code 2.1** An example to illustrate some of the Tidyverse paradigms

```r
1  demo <- penguins[1, c("bill_length_mm", "bill_depth_mm", "sex")]
2
3  labelled <- function(x, unit = NULL) {
4    vctrs::new_vctr(x, class = c("label", class(x)), unit = unit)
5  }
6  vec_ptype_abbr.label <- function(x) attr(x, "unit")
7
8  add_unit <- function(.data, x, unit, transform = ~.) {
9    xq <- rlang::as_string(rlang::enexpr(x))
10   f <- rlang::as_function(transform)
11   dplyr::mutate(.data, !!xq := labelled(f(.data[[xq]]), .env$unit))
12 }
13
14 add_unit(demo, bill_depth_mm, "cm", transform = ~ . / 100)

   # A tibble: 1 x 3
     bill_length_mm bill_depth_mm sex
             <dbl> <cm>          <fct>
   1          39.1 0.187         male
```

## 2.2 Syntax for data visualisation

Trellis displays, also known as small multiples or faceted plots, are a widely used and effective method in data visualization. These displays involve arranging multiple plots, each representing the same plot type but with different subsets of data, in a structured manner (Becker, Cleveland, and Shyu 1996). Constructing trellis displays requires a mechanism for arranging the plots according to their aspect ratios, margins, and panel dimensions (Murrell 1999). Typically, trellis displays are configured in a rectangular layout, with each plot panel having the same rectangular dimensions. Consequently, a high-level specification is often employed to define the display layout, such as the number of rows and columns or the mapping of plots or data subsets to individual panels.

The layouts in `ggplot2` (Wickham 2010) are primarily controlled by `Facet` objects. More specifically, `ggplot2` composes plots based on assembling modular objects (data, aesthetic mapping, layer, geometric objects, statistics, coordinate system, theme, facet, scales, and guides). These modular objects are primarily defined though an object oriented system called `ggproto`, inspired by the grammar of graphics by Wilkinson (2005). A new plot in `ggplot2` is made by substituting these modular objects.

Code 2.2 demonstrates a ggplot2 syntax for constructing a trellis display using the `penguins` data. Lines 2-3 specifies the primary data and the aesthetic mapping of data to graphical elements. Line 4 specifies the plot type (with default statistics calculated from frequency of bin width of 1 and bars outlined in black). Line 5 details the layout of the trellis display (referred to as facet in `ggplot2`), indicating that the panel data is separated by species, organized into one column, with the row order adjusted so that the first factor level appears at the bottom and the last factor level appears at the top.

Code 2.2 (Tidyverse approach): data visualisation

```
library(ggplot2)
ggplot(penguins) +
  aes(x = bill_length_mm, fill = sex) +
  geom_histogram(binwidth = 1, color = "black") +
  facet_wrap(~species, ncol = 1, as.table = FALSE)
```

```
Warning: Removed 2 rows containing non-finite outside the scale range
(`stat_bin()`).
```

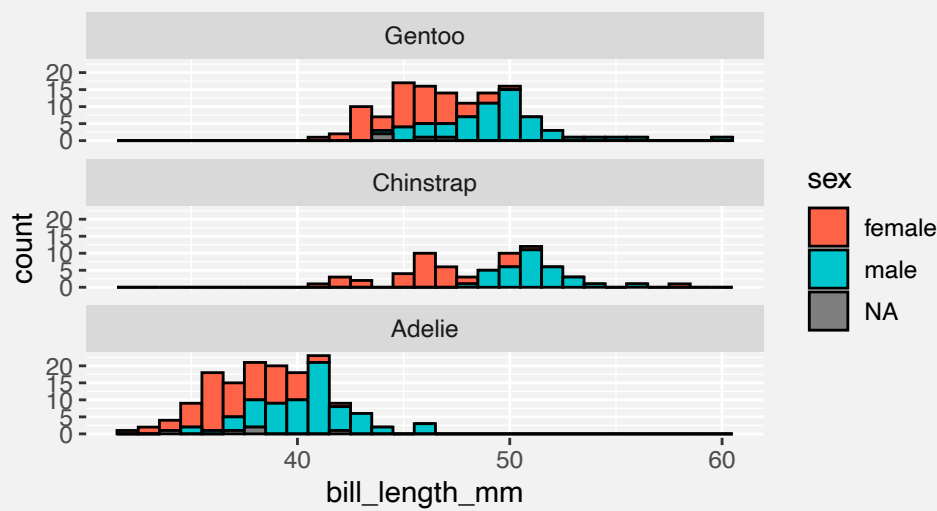

Base R (through the `graphics` package) specifies the layout of plots using `par()`, after which users print each individual plot to the graphics output sequentially. Code 2.3 shows a near equivalent output to Code 2.2 using a Base R approach. Specifically, line 1 sets the layout, line 2 calculates the minimum and maximum values of the bill length to establish a common limit for the $x$-axis, and line 5-15 subsets the data by species and sex and plot the histogram for the corresponding data subset. Then the lines 13 and 20-22 adds the legend and labels. Some of the code (lines 9-12 and 16-17) is used to modify the default aesthetics for a more cohesive display. In this instance, data points that have missing sex values are silently dropped.

Code 2.3 (Base R approach): data visualisation

```r
par(mfrow = c(3, 1), mar = c(0.5, 4, 2, 0.2), oma = c(4, 0, 0, 0))
xlims <- range(penguins$bill_length_mm, na.rm = TRUE)
for (aspecies in c("Gentoo", "Chinstrap", "Adelie")) {
  for (asex in c("female", "male")) {
    hist(
      with(penguins, bill_length_mm[sex == asex & species == aspecies]),
      col = ifelse(asex == "female", "tomato", "turquoise3"),
      breaks = seq(xlims[1], xlims[2] + 0.5, by = 1),
      xlim = xlims,
      ylim = c(0, 20),
      axes = FALSE,
      xlab = NULL,
      main = aspecies,
      add = asex == "male"
    )
    axis(2, labels = TRUE, tick = TRUE)
    axis(1, labels = FALSE, tick = FALSE)
  }
}
title(xlab = "Bill length (mm)", outer = TRUE)
axis(1, labels = TRUE, tick = TRUE)
legend(55, 20, legend = c("Female", "Male"), fill = c("tomato", "turquoise3"))
```

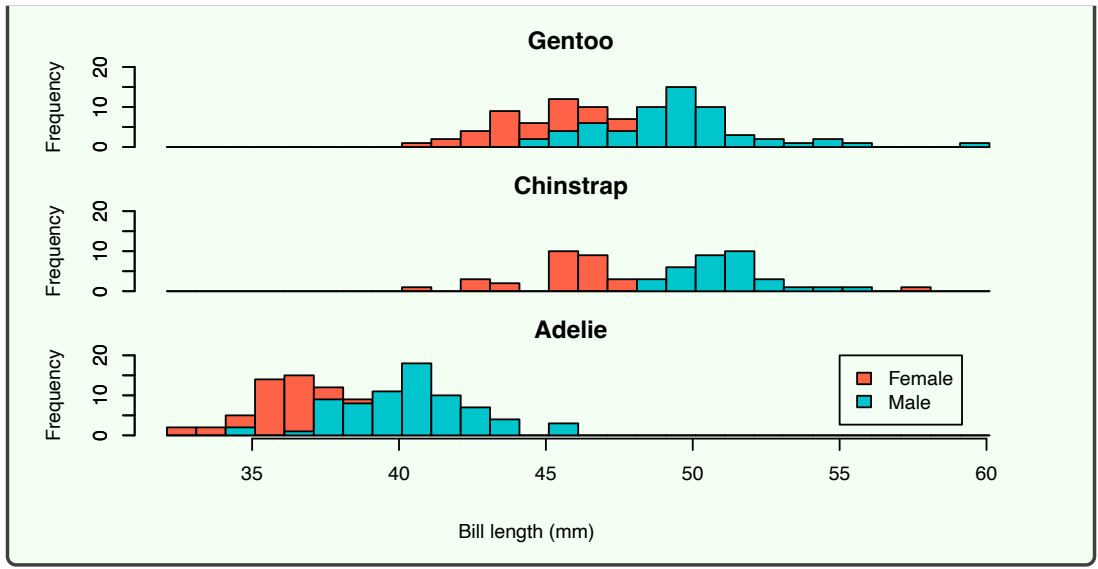

The Base R approach to creating a trellis display is evidently cumbersome, requiring users to manually construct individual plots and modify default aesthetics. A notable package that simplified the creation of trellis plots in R is the `lattice` package (Sarkar 2008).

The `lattice` package, along with its extension `latticeExtra` (Sarkar and Andrews 2022), introduced a formula syntax for specifying the layout for trellis displays. A typical formula syntax had the form "yvar ~ xvar | cvar" where yvar, xvar and cvar represent the optional data variables for the $y$-axis, $x$-axis and conditioning variable for data splitting, respectively. The equivalent `lattice` syntax for the output in Code 2.3 is presented in Code 2.4. Line 3 designates the variables from the data that maps to the $x$-axis and the conditioning variable. The left hand side of the formula is omitted as the $y$-axis is computed as the count (line 6) from the given bin widths in line 5 (to keep the same bin width as Code 2.3). Lines 8-13 writes the panel function to superimpose the histograms by different groups (specified in line 7). Then line 15 specifies the layout such that there is one column and three rows (the default display would be one row and three columns). Finally, the line 16 creates a legend for the color of the histograms. This code is more verbose than a typical `lattice` code because creating an overlapping histogram necessitates overwriting the panel function.

````
Code 2.4 (lattice approach): data visualisation
1  library(lattice)
2  histogram(
3    ~ bill_length_mm | species,
4    data = penguins,
5    breaks = seq(xlims[1], xlims[2] + 0.5, by = 1),
6    type = "count",
7    groups = sex,
8    panel = function(...) {
9      panel.superpose(
10       ...,
11       panel.groups = panel.histogram,
12       col = c("tomato", "turquoise3")
13     )
14   },
15   layout = c(1, 3),
16   auto.key = list(rectangles = FALSE, col = c("tomato", "turquoise3"))
17 )
````

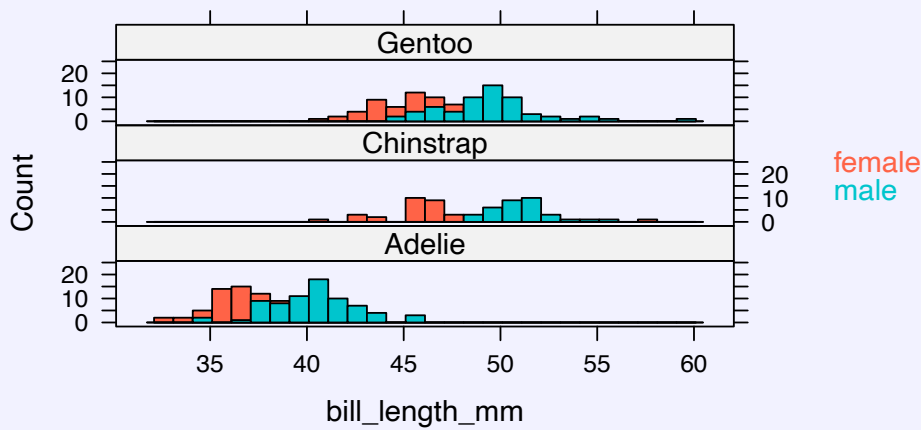

The usage of `lattice` have, however, largely been eclipsed by `ggplot2` (see Supplementary materials for the download statistics). Evidently, `ggplot2` approach in Code 2.2 is more concise and intuitive than the Base R and `lattice` approaches in Codes 2.3 and 2.4, respectively. As seen in the above examples, both Base R and `lattice` approaches generally require some manual bookkeeping for complex graphics, e.g. specification of the colors for the legend, which invites possibility of human error. In addition, both Base R and `lattice` silently dropped data points with missing values in sex and bill length, where as `ggplot2` either provides a warning or plot the missing value as another category as the default.

There are examples where Base R and `lattice` approaches are more concise and intuitive than `ggplot2`, but these are often limited to simpler graphs, e.g. `hist(penguins$bill_length_mm)` for histogram in Base R and `ggplot(penguins) + geom_histogram(aes(bill_length_mm))` for `ggplot2` where the latter requires more explanation to new users. However, the code complexity in `ggplot2` do not increase as greatly as in Base R when mapping additional variables to the plot. This makes `ggplot2` especially user-friendly and efficient for visualising more complex datasets.

## 2.3 Syntax for data wrangling

Structured Query Language (SQL) is an industry standard for performing data operations on relational databases (Melton 1998). The syntax of SQL consists of a series of English-like statements with each statement in a new line. This syntax is reminiscent of (and possibly what inspired) data wrangling using `dplyr` in conjunction with pipes.

For example, Code 2.5 demonstrates the application of the five major functions in `dplyr` (`filter`, `select`, `mutate`, `summarise` and `arrange`) to manipulate the `penguins` data. First, the data is filtered to retain only observations with a non-missing sex classification (line 3), followed by the selection of columns named species, sex, island and those beginning with "bill" (line 4). Next, a new column named `bill_area_mm`, calculated by multiplying `bill_length_mm` by `bill_depth_mm`, is inserted (line 5). Subsequently, the first quartile is obtained for all numeric variables by species and sex (line 6-9). Finally, the data is arranged by species and sex (line 10).

Code 2.5 (Tidyverse approach): data manipulation

```r
library(dplyr)
penguins |>
  filter(!is.na(sex)) |>
  select(species, sex, island, starts_with("bill")) |>
  mutate(bill_area_mm = bill_length_mm * bill_depth_mm) |>
  summarise(
    across(where(is.numeric), ~ quantile(., 0.25)),
    .by = c(species, sex)
  ) |>
  arrange(species, sex)

# A tibble: 6 x 5
  species    sex     bill_length_mm bill_depth_mm bill_area_mm
  <fct>      <fct>            <dbl>         <dbl>        <dbl>
1 Adelie     female            35.9            17          623
```

```
2 Adelie    male                39         18.5       730.
3 Chinstrap female              45.4       17         777.
4 Chinstrap male                50.0       18.8       947.
5 Gentoo    female              43.8       13.8       618.
6 Gentoo    male                48.1       15.2       728.
```

As noted by Çetinkaya-Rundel et al. (2022), the readability of code and its transferability to and from SQL are significant advantages of using dplyr. Although from personal experience, data wrangling in Base R is typically characterized by the use of $ or [ operators, equivalent functions are available, as demonstrated in Code 2.6. Lines 2, 3-4, 5, 7 and 8-10 in Code 2.6 are equivalent to lines 3, 4, 5, 6-9 and 10 in Code 2.5, respectively. However, arrange() is replaced by a combination of anonymous function, the operator [ and order() function, and there is no easy way to extract the first quartile for numeric variables only (line 7). From personal observation, pipes are less commonly used by Base R users. Native pipe (|>) was only introduced from R version 4.1.0.

Code 2.6 (Base R approach): data manipulation

```r
penguins |>
  subset(!is.na(sex)) |>
  subset(select = c(species, sex, island,
                    grep("^bill", colnames(penguins)))) |>
  transform(bill_area_mm = bill_length_mm * bill_depth_mm) |>
  # no easy way for numeric only
  aggregate(function(x) quantile(x, 0.25), by = . ~ species + sex) |>
  {
    \(d) d[order(d$species, d$sex), ]
  }()
```

```
    species    sex island bill_length_mm bill_depth_mm bill_area_mm
1    Adelie female      1         35.900          17.0      623.000
4    Adelie   male      1         39.000          18.5      730.480
2 Chinstrap female      2         45.425          17.0      777.310
5 Chinstrap   male      2         50.050          18.8      946.695
3    Gentoo female      1         43.850          13.8      618.180
6    Gentoo   male      1         48.100          15.2      728.460
```

There are a few notable differences between the Tidyverse and Base R approaches:

- Tidyverse employs selection helpers (in the tidyselect package) for column selection, as seen with the starts_with() function in line 4 of Code 2.5. These selection helpers function only within packages that implement `tidyselect` in the backend, thus Base R approach requires the use of `grep()` in place of `starts_with()`, as shown in line 4 of Code 2.6. There is a `startsWith()` function in Base R, however it only returns a logical vector and the `select` argument in `subset()` cannot combine a logical vector with other selections (i.e. numeric, character or unquoted name) so cannot be used in this instance.

- Group operations via `summarise()` in Tidyverse and `aggregate()` in Base R are specified differently. The latter employs a formula syntax (line 7 of Code 2.6) to define the grouping structure, while the former uses `tidyselect` for variable selection (line 8 of Code 2.5) or can be specified as a separate process in the pipeline by using `group_by()`.

- Tidyverse employs a `purrr`-like formula syntax (line 7 of Code 2.5) as a shorthand for anonymous functions. Note that Base R also introduced a shorthand syntax for anonymous functions from R version 4.1.0 (see lines 8-10 of Code 2.6).

Another data wrangling approach that has gained traction for its performance with large data is `data.table` (Barrett et al. 2024). The syntax of `data.table` uses square bracket operator `[` to often make in-place modifications to the data, which is more efficient than the copy-on-modify behaviour of `dplyr`. While using pipes is an atypical approach for `data.table`, we present the usage with pipes in Code 2.7 to encourage direct comparison between the syntaxes. Lines 3, 4-7,8, 9-13 and 14 in Code 2.7 is the `data.table` equivalent to lines 3, 4, 5, 6-9 and 10 in Code 2.5, respectively. Notably, `data.table` uses a reserved symbol `.SD` to refer to the subset of data that is not part of the grouping structure and `.SDcols` to specify the columns to be operated on.

````
Code 2.7 (data.table approach): data manipulation
```
1 library(data.table)
2 as.data.table(penguins) |>
3   _[!is.na(sex)] |>
4   _[,
5     c(.(sex = sex, species = species, island = island), .SD),
6     .SDcols = patterns("^bill")
7   ] |>
8   _[, bill_area_mm := bill_length_mm * bill_depth_mm] |>
9   _[,
10    lapply(.SD, \(x) quantile(x, 0.25)),
11    .SDcols = is.numeric,
12    by = .(species, sex)
13  ] |>
14  _[order(species, sex), ]

     species     sex bill_length_mm bill_depth_mm bill_area_mm
       <fctr>  <fctr>          <num>         <num>        <num>
1:    Adelie  female         35.900          17.0      623.000
2:    Adelie    male         39.000          18.5      730.480
3: Chinstrap  female         45.425          17.0      777.310
4: Chinstrap    male         50.050          18.8      946.695
5:    Gentoo  female         43.850          13.8      618.180
6:    Gentoo    male         48.100          15.2      728.460
```

While the pipeline syntax for Tidyverse and Base R approaches in Code 2.5 and Code 2.6, respectively, are similar, there are still frictions in the Base R approach that perhaps discourage the use of the pipes in Base R. First, there is a need to re-specify the data object in the argument for column selection in line 4 of Code 2.6, necessitating the creation of intermediate objects if the data was modified. Secondly, `aggregate()` nor `subset()` include a method that selects columns based on a mix of predicate functions and its names, again encouraging creation of an intermediate character or logical vector for selection. Thirdly, the ordering of the rows requires an anonymous function (lines 8-10 of Code 2.6) if using a pipeline approach.

The syntax for `data.table` present a fundamentally different syntax to `dplyr` by performing operations within [. While `data.table` has superior performance than `dplyr`, the `dplyr` package has far more download than `data.table` (see Supplementary materials). This suggests that the interface design may be the driving factor for the preference of `dplyr` over `data.table`.



The R language and the contributed packages have evolved over time, mutually shaping each other's development. Tidyverse, for instance, inspired the introduction of pipes and lambda functions to Base R (Çetinkaya-Rundel et al. 2022). Similarly, `.by` argument in `dplyr` functions was introduced from version 1.1.0 taking inspiration from the inline by argument from `data.table` (in news of Wickham et al. 2023).

# 3 Approaches to user interface designs

The principles and guidelines pertaining to user interface (UI) designs have limited discussions within the field of statistics; however, they are extensively explored in disciplines such as human-computer interaction, human factors, and cognitive ergonomics, where human interaction is a primary concern in the development of products or environments. Ruiz, Serral, and Snoeck (2021) empirically synthesised UI design principles mentioned in literature to 36 UI design principles. As a number of principles had a similar theme (e.g. "offer informative feedback" and "good error messages"), we further distilled these principles (with the assistance of a large language model) into four overarching themes as presented in Table 2.

Table 2: The four overarching themes for UI design principles in literature.

| Theme | Description | UI design principles |
|---|---|---|
| 1 | Promote user-centered design and iterative improvement | Use real-world metaphors (transfer); Provide a good conceptual model of the system; Know the user; Empirical measurement; Iterative design to remove usability problems; Understand the tasks; Reuse; Accommodate users with different skill levels; Integrated design |
| 2 | Provide a clear and intuitive user experience | Offer informative feedback; Strive for consistency; Minimize user's memory load; Simple and natural dialog; Speak the user's language; Help and documentation; Make things visible; Affordance; Design dialogues to yield closure; Recognition rather than recall; Flexibility and efficiency of use; Structure the user's interface; Provide visual cues; Display inertia |



| Theme | Description | UI design principles |
|---|---|---|
| 3 | Foster user autonomy and error prevention | Prevent errors; Good error messages; Provide shortcuts; Provide clearly marked exits; Actions should be reversible; Support internal locus of control; Constraints; Give the user control; Help users recognize, diagnose and recover from errors; Allow users to customize the interface |
| 4 | Encourage exploration and accessibility | Cater to universal usability; Allow users to change focus; Encourage exploration |

In the realm of UI design, two primary approaches are broadly identified: *user-centered design* (UCD) and *ecological interface design* (EID) (Wu et al. 2016). These approaches focus on opposite ends of the spectrum ("work domain" and end-user) as illustrated in Figure 1. In the context of EID, "work domain" refers to the system or environment within which work is performed. Detailed discussions on UCD and EID are provided in Sections 3.1 and 3.2, respectively.

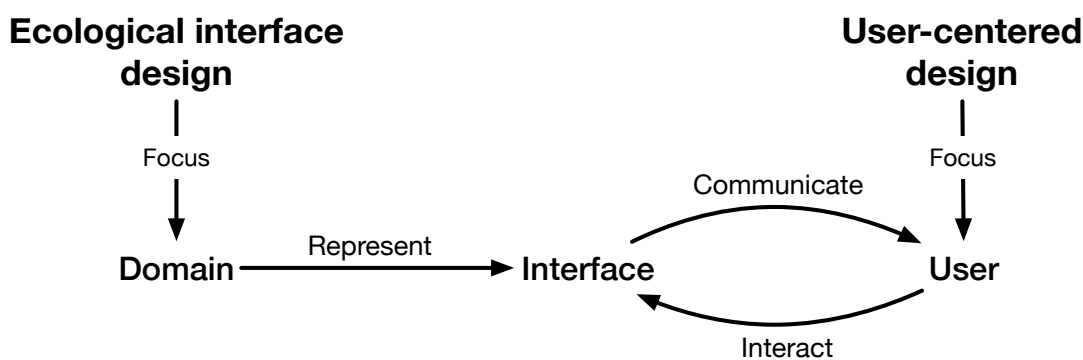

Figure 1: An interface embodies elements of the work domain that convey to the user the available functionalities and parameters for interaction. Ecological interface design emphasizes making the constraints and relationships within the work domain perceptually apparent to the user. In contrast, user-centered design prioritizes human factors in its design approach, primarily relying on user feedback.



## 3.1 Human/User-centered design

The International Organization for Standardization defines human-centered design (HCD) in the ISO 9241 series as the "approach to systems design and development that aims to make interactive systems more usable by focusing on the use of the system and applying human factors/ergonomics and usability knowledge and techniques" (ISO 9241-210 2010). As noted by ISO 9241-210 (2010), UCD is used interchangeably with HCD in practice; however, HCD also addresses the impacts on stakeholders beyond just the end users.

A notable element of HCD/UCD is the active involvement of users in the design process. According to ISO 9241-210 (2010), a human-centered approach should adhere to the following principles:

(a) the design is based upon an explicit understanding of users, tasks and environment,

(b) users are involved throughout design and development,

(c) the design is driven and refined by user-centered evaluation,

(d) the process is iterative,

(e) the design addresses the whole user experience, and

(f) the design team includes multidisciplinary skills and perspectives.

However, not all agree with these published principles (for an extensive discussion, see Chammas, Quaresma, and Mont'Alvão 2015).

In a seminar paper by Gould and Lewis (1985), it is recommended that a HCD/UCD starts with user interviews and discussions *prior to* system design. They also advocate for an empirical assessment of the system in terms of user interaction and iterative improvement through repeated design cycles. Notably, these recommendations are not common practice in the design of statistical software.

## 3.2 Ecological interface design

HCD/UCD is based on the premise that users are best positioned to determine what works for them (or other users) for achieving specific functional purposes. However, this assumption does not always hold in the context of complex systems, where users may not fully understand the system (Burns and Hajdukiewicz 2004). In such cases, ecological interface design (EID) can assist users in learning about the system as they use it, and managing unexpected situations (Rasmussen and Vicente 1989; Burns and Hajdukiewicz 2004).



EID has been successfully applied in a broad range of sociotechnical systems (e.g. power distribution, transportation, military, medicine, and network management) for over 30 years (Bennett and Flach 2019). Initially proposed by Rasmussen and Vicente (1989), EID aims to mitigate human errors by designing systems that account for the three categories of human behaviour, known as the *SRK taxonomy*: skill-based (autonomous actions), rule-based (following if-else decision tree type action), and knowledge-based (reasoning within the work domain).

EID addresses a wide range of human behaviours by concentrating on the "work domain", i.e. the system or environment, defined in terms of the system's functional purposes and constraints, rather than the user's actions, and make these system properties visible to the user. E.g., in `ggplot2`, the work domain is the grammar of graphics as adapted to the R language. Wilkinson (2005) specified a plot through six functional purposes: a set of data operations, variable transformations, scale transformations, a coordinate system, geometric objects and their aesthetic attributes, and guides. Building on this, Wickham (2010) adapted these functional purposes accessible to users within the constraints of the R language. Clearly, the primary development of `ggplot2` is closer to EID.

The central idea of EID is to organise and make visible the system constraints and relationships within the work domain to users, effectively making the invisible visible (Rasmussen and Vicente 1989). This approach allows users to make decisions using either their perceptual systems – with minimal cognitive processing – or through analytical reasoning, which requires more intensive cognitive processing by leveraging finer details and their knowledge (Vicente and Rasmussen 1992). This concept is derived from Gibson's theory of direct visual perception (Gibson 2015), which suggests that observers can directly perceive meaningful information from their environment, leading to the notion of *affordance*, where the attributes of an object naturally indicate how it can be used.

To apply EID in practice, a thorough analysis of the work domain is necessary, and the scope or system boundary must be established prior to the interface design (Burns and Hajdukiewicz 2004). The developer or designer is expected to possess a comprehensive understanding of the work domain. Defining the scope involves determining the extent of user control, identifying essential elements of the work domain, and ascertaining what information should be visible to the user. Burns and Hajdukiewicz (2004) recommends beginning with defining the functional purposes of the system, followed by defining the physical form (in the context of statistical software, what the users can see, e.g. such as the naming and structure of functions, arguments, and objects), and then filling the gaps (i.e. connecting the syntax with functional purpose). By ensuring that all functional purposes are covered by the syntax, users have means to act for changing the system output.



# 4 Lessons learnt from the Tidyverse interface design

The design principles of the Tidyverse (Wickham 2024) generally aligns with the themes outlined in Table 2, potentially explaining why it has resonated well with the masses. Specifically, the human-centered approach aligns with Theme 1; consistency in syntax corresponds to Theme 2; composablity, which empowers users to explore by combining modular functions, aligns approximately with Themes 3 and 4; and inclusivity align with the emphasis on accessibility in Theme 4. Çetinkaya-Rundel et al. (2022) supports the notion that the Tidyverse has been developed following a user-centered design process (Theme 1), citing the evolution of reshaping data (which included a short user survey) as a prime example of this process. Overall, the Tidyverse team, whether intentionally or not, largely follow best practices in UI design.

Classical recommendations for HCD insist on direct engagement with users *prior to* system design and empirical assessment of user interactions with the system. These practices are largely absent in the development of the Tidyverse (and other statistical systems), although the Tidyverse team do actively solicits feedback post system design (Çetinkaya-Rundel et al. 2022). The extent to which user feedback is incorporated remains at the developers' discretion, so developers must formulate an internal mechanism to decide which user feedback to incorporate. Wickham (2024) claim this mechanism is based on the design principles: human-centered, consistency, composability and inclusivity, however, this is debatable as explained next.

The opposite of HCD is to design without user input or focus on the end-user. Although this seems contrary to the approach of the Tidyverse design, it warrants consideration given the success of EID for complex systems (see Section 3.2). A prime example where user feedback may not have helped in its initial development is `ggplot2`. In `ggplot2`, a plot is assembled from graphical objects pertaining to different elements (geometric object, statistics, scales, guides, data, aesthetic mapping, and so on), and functions are named to correspond with these elements (e.g. functions that prefix with `scale_` control the scales). This structure allows for greater expressive capability in plotting with a limited number of functions but introduces a steeper learning curve. Users (particularly those used to using a less verbose, recipe-like approach to creating plots in Base R) may initially find the `ggplot2` syntax challenging. To address this challenge, `ggplot2` historically offered `qplot()` for drawing quick plots, but it has been deprecated in favour of using the complete `ggplot()` syntax. This observation suggests that a user's desire to have a recipe-like plotting system via `qplot()` was not helpful. Despite its complexity, `ggplot2` remains valuable, consistently ranking among the top 10 most downloaded R packages for over a decade (see Supplementary materials). EID works better than UCD for a complex system when the user does not know the best. Arguably, `ggplot2` is successful



because its design approach is similar to EID, i.e. focus on the perceptual visibility of functionalities in the plotting system.

Most `ggplot2` users likely do not comprehend the entire system. From my own experience, I initially learned to use `ggplot2` by adapting numerous code snippets copied and pasted from the Internet (akin to a rule-based behaviour in the SRK taxonomy). Gradually, I discerned the pattern in the syntax structure related to the graphical output (this would have been all obvious have I read Wickham (2016), but I did not until much later). My proficiency in specifying plots with `ggplot2` developed gradually through frequent use to make various plots. However, to truly create any plot, knowledge of the internal system (as well as related systems like `gtable` and `grid`) is essential. `ggplot2` offers external methods for extending the system by replacing the `ggproto` objects (`Coord`, `Facet` `Geom`, `Guide`, `Layout`, `Position`, `Scale`, and `Stat`), although this knowledge is typically beyond that of most users, who instead rely on extension packages. An example of this is `ggpattern` (FC, Davis, and ggplot2 authors 2025) where users can specify a pattern to fill geometric areas based on a data variable. To achieve a similar result in `ggplot2` would be extraordinary difficult for typical users without using `ggpattern`. The development of `ggpattern` is only possible from understanding the `ggplot2` internal.

The interface design of `ggplot2` with some respect address each of the behaviour in the SRK taxonomy by providing visibility of the relationship between components (see Figure 2). Some packages purported to be `ggplot2` extension packages are wrappers to make a complete plot (much like `qplot()`) rather than extensions and are inconsistent with the original interface design, therefore these would not be considered addressing the SRK taxonomy.

The central premise of EID is that system design can aid users in navigating unexpected situations by making them aware of the system's constraints and relationships. Çetinkaya-Rundel et al. (2022) highlights how the Tidyverse improves error recovery and prevention by, for example in `ggplot2`, providing warnings when observations with missing values are dropped.

Even in data wrangling, users may be unaware of how the system works, and it can be beneficial to purposefully prioritise the visibility of system constraints and relationships. For example, some users (myself included) found the use of the selection helper `where()` such that it results in the nesting of three functions (`summarise(across(where()))`) like in line 7 of Code 2.5 visibly different in style and initially disliked it. If this feedback is taken on, then it will call for a removal of `where()`. While the system can function without `where()`, it issues a warning, as seen in the simpler example in Code 4.1, when selecting all columns that are factors. If the system functions correctly, why is the warning necessary? This explanation follows.



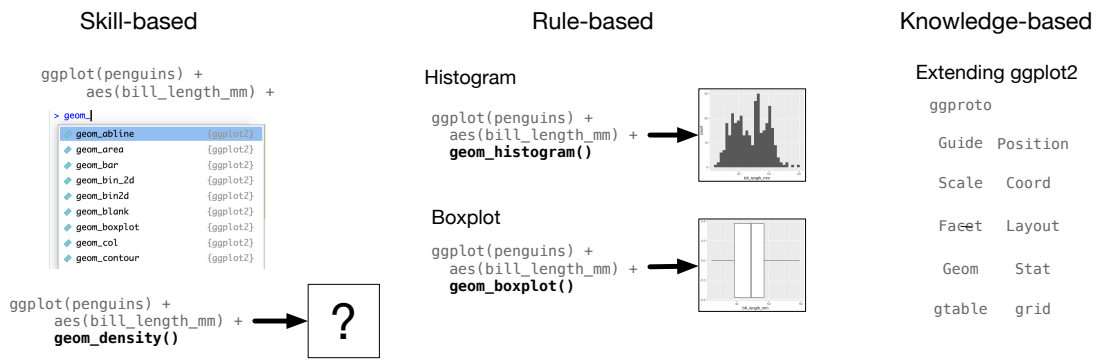

Figure 2: The objective of creating a plot can be accomplished through various modes of behavior: skill-based behaviour, where the user is familiar with the syntax "grammar"; rule-based behaviour, involving the adaptation of copied code snippets; and knowledge-based behaviour, where the user can reason how the internal system is assembling the graphical objects and modifies components using their understanding. In ggplot2, the graphical elements are exposed in the syntax, and upon rendering the graphics, the user is informed of the current state. This setup offers the user an opportunity to deduce how the system works.

Code 4.1

```
1 penguins |> select(is.factor) |> head(2)

Warning: Use of bare predicate functions was deprecated in tidyselect 1.1.0.
i Please use wrap predicates in `where()` instead.
  # Was:
  data %>% select(is.factor)

  # Now:
  data %>% select(where(is.factor))

# A tibble: 2 x 3
  species island    sex
  <fct>   <fct>     <fct>
1 Adelie  Torgersen male
2 Adelie  Torgersen female
```

The syntax `where()` is used to *unequivocally convey the intention of the user* to use a predicate function. Code 4.2 presents an example of an unintended consequence when there is a column named `is.factor`. In such a case, column selection in line 2 results in



selecting this column itself rather than invoking the predicate function `is.factor()`.

Code 4.2
```
1 penguins_demo <- penguins |> mutate(is.factor = NA)
2 penguins_demo |> select(is.factor) |> head(2)

  # A tibble: 2 x 1
    is.factor
    <lgl>
  1 NA
  2 NA
```

The use of `where()`, as shown in Code 4.3, is a deliberate design choice aimed at ensuring both the system and the user understand that the input is intended to be a predicate function. Similar reasoning underlies the use of the selection helpers `all_of()` and `any_of()` in `tidyselect`, which are used by `dplyr` for column selection.

Code 4.3
```
1 penguins_demo |> select(where(is.factor)) |> head(2)

  # A tibble: 2 x 3
    species island    sex
    <fct>   <fct>     <fct>
  1 Adelie  Torgersen male
  2 Adelie  Torgersen female
```

Overall, the development of Tidyverse appear to involve a mix of UCD and EID. The Tidyverse team are responsive to the large volume of feedback they receive (e.g. social media, GitHub issues, Q&A platforms, etc), however, examining `tidyverse` suggests that there is focus on the work domain. This means that the interface should be designed to reveal how the system works, including its limit.

## 5 Discussion

The Tidyverse has gained significant popularity and influence in the practice and development of other statistical software that adopt a "tidy approach" (see Section 2.1). This influence suggests that there are factors contributing to its wide-spread adoption, with interface design being a potential significant factor. We aimed to isolate the design



approaches within the Tidyverse that could aid in the effective design of statistical software beyond its context. We compared the Tidyverse syntax with Base R and alternative approaches (Section 2) and examined the Tidyverse design (Section 4) in relation to a summary of best practice for user interface (UI) design, and two contrasting UI design approaches: human (or user) centered design (HCD) and ecological interface design (EID) (Section 3). We generally find that Tidyverse design adheres to best practice in UI design and employs a mix of HCD and EID approaches.

On the other hand, the R Core Team that develop Base R focus on backward compatibility, and therefore are cautious to make iterative changes to the interface design. There are justified reasons for this, but this results in a less consistent interface design due to historical development, in contrast to the advice of iterative improvement in Table 2. Additionally, alternative approaches to the Tidyverse appear to not extend the interface design to the full work domain. For example, the complexity in specifying an overlapping histogram in `lattice` (Code 2.4) suggests that it may not have been in consideration for the design. Similarly, the lack of a function to order rows for tabular data where the first argument is tabular data (Code 2.6) suggest that the data wrangling in Base R was not designed for working fully with pipes. The lack of consideration of the full work domain gives rise to inconsistent user experience in contrary to the advice in Table 2. Therefore, we recommend that analysis of the full work domain (including inputs and outputs of the system) is conducted prior to the interface design. Then developers should ensure that the interface has complete coverage of the outputs by mapping each syntax to a functional purpose. The syntax must cover all the functional purposes (i.e. the mapping must be surjective). This surjective mapping promotes perceptual visibility of system constraints and relationships. In another words, the user builds a cognitive awareness of the syntax and its effect within the system.

Interface design serves as a form of communication. A better interface design can be thought of as optimising the communication of intentions of the developers and the users. The Tidyverse team often seeks qualitative feedback upon software release, which would have aided in their design process, but as discussed in Section 4, the extent the user feedback informs the design is unclear. Elements of its design suggest that there are aspects of the system design that prioritise visibility of system constraints to help users cope with unexpected situations. Therefore, the design and development of a statistical software is likely to benefit from both HCD and EID approaches.

While this paper posits that the success of the Tidyverse may derive from its interface design, other factors should not be discounted. One such factor is documentation. Well-documented functions with relevant examples are sure to be helpful to the users and may promote higher usage. The Tidyverse boasts extensive documentation and a large user base that generates numerous helpful examples and answers on forums. Additionally, the Tidyverse is primarily created and maintained by employees of Posit PBC (formerly



RStudio) who possess greater resources at hand to develop and promote the Tidyverse. As such, another reason for its popularity may be the professionalism and advocacy behind the team. These factors are hard to emulate for individuals, groups or organisations with constraint resources, as is the case for most academic researchers who propose new statistical methodologies.

Naturally, a better interface design is made in mind for the average or mode target user group. A suitable interface for the average user does not necessary mean that it will suit everyone in the group. While the Tidyverse is popular, this does not mean it fits everyone and there are active opponents to its use, leading to contentious debate regarding the teaching of Base R and Tidyverse (N. Matloff 2023). Its popularity suggests it resonates well with a broad audience, but the lack of fit is expected for some individuals. However, various metrics (downloads, citations, usage, and so on) show a substantial user base impacted positively by the Tidyverse, which warrants thoughtful consideration before dismissal.

Arguably, the most successful aspect of the Tidyverse design is its engagement with the user community. The Tidyverse has likely introduced more users to R, made R more accessible to a diverse audience, and inspired the development of statistical software adopting similar approaches. Similar to Burns and Proulx (2002)'s attempt to influence a social problem (i.e. gambling) through interface design, an effectively designed interface for statistical software can significantly impact the community and the practice of statistics on a large scale.

Statistical modelling is another prominent task in data analysis that has not been discussed in this paper. The `tidymodels` (Kuhn and Wickham 2020) make some strides in the interface design for employing Tidyverse principles to modelling and machine learning. Future research could benefit from further discussion in the interface design of statistical modelling and other data analysis tasks, and the development of community guidelines for the design of statistical software.

# Computational Details

This paper is written using Quarto (Allaire et al. 2024) version 1.8.24 powered by Pandoc version 3.6.3 (MacFarlane, Krewinkel, and Rosenthal 2024) and pdfTeX version 3.141592653-2.6-1.40.26 (TeX Live 2024). The diagrams are drawn using OmniGraffle version 7.24.1 (v205.48.5). The code was formatted using Air version 0.7.1.



# Acknowledgement

I thank Francis Hui for his feedback on the initial draft of this manuscript that has helped to improve this manuscript. I also thank the three anonymous reviewers for their constructive feedback that has significantly improved the manuscript.